\documentclass[twocolumn,pre,aps,showpacs,superscriptaddress]{revtex4}

\usepackage[english]{babel}
\usepackage{epsfig}
\usepackage{amsmath}
\usepackage{graphicx}

\begin{document}

\title{Lagrangian Particle Statistics in Turbulent Flows from a Simple Vortex Model}

\author{M.~Wilczek}
\email{mwilczek@uni-muenster.de}
\affiliation{Institute for Theoretical Physics, University of M\"unster,
Wilhelm-Klemm-Str. 9, D-48149 M\"unster, Germany}
\author{F.~Jenko}
\affiliation{Max-Planck-Institut f\"ur Plasmaphysik,
Boltzmannstr. 2, D-85748 Garching, Germany}
\author{R.~Friedrich}
\affiliation{Institute for Theoretical Physics, University of M\"unster,
Wilhelm-Klemm-Str. 9, D-48149 M\"unster, Germany}
\email{mwilczek@uni-muenster.de}

\date{\today}

\pacs{47.10.ad,47.27.-i,47.27.E-,02.50.Fz}


\begin{abstract}

The statistics of Lagrangian particles in turbulent flows is considered in
the framework of a simple vortex model. Here, the turbulent velocity field
is represented by a temporal sequence of Burgers vortices of different
circulation, strain, and orientation. Based on suitable assumptions about
the vortices' statistical properties, the statistics of the velocity increments
is derived. In particular, the origin and nature of small-scale intermittency in
this model is investigated both numerically and analytically.

\end{abstract}

\maketitle

\section{Introduction}

Our understanding of the spatio-temporal properties
of turbulent flows is still fragmentary \cite{Frisch95book,Pope00book, davidson04book}. In recent years, 
however, significant progress could be achieved by focusing on the dynamics of Lagrangian particles (see, e.g., Ref.~\cite{yeung02arf} and references therein). 
Especially laboratory experiments have given useful insight to important Lagrangian statistical properties.
For example, the statistics of Lagrangian velocity
increments \cite{mordant01prl,mordant04njp} and those of the turbulent acceleration of a 
fluid particle \cite{laporta01nat, mordant01prl} have been measured. Moreover, direct numerical
simulations on large-scale computers have been performed \cite{biferale04arx} which
allow to study particle trajectories and their statistical properties from yet
another point of view. While all results obtained from laboratory and computer
experiments seem to be consistent so far, the underlying physical mechanisms
are still subject to discussion. In particular, it has not yet been possible
to arrive at a theoretical derivation of the single particle velocity increment
distribution.  

There are hints, though, that certain turbulent structures are largely responsible
for the statistical properties of Lagrangian particles. It has been shown by direct
numerical simulations \cite{jimenez93jfm, farge01prl} and by thorough experimental investigations
\cite{mouri03pre} that the turbulent field contains elongated vortex filaments which
can be interpreted as Burgers vortices. The latter are well-known solutions of the
Navier-Stokes equation \cite{burgers48aam}. In fact, these vortices have been called the
{\em sinews of turbulence} by Moffat and coworkers \cite{moffat94jfm}. Hatakeyama and Kambe recently modelled turbulent fields by a random arrangement of such Burgers
vortices \cite{hatakeyama97prl}. Interestingly, through this procedure they were able to reproduce the
multifractal scaling behavior of the longitudinal structure functions $S_n(r)$.

In the present paper, we shall employ a similar approach. More precisely, we will
consider the Lagrangian statistics of a tracer particle in a turbulent velocity
field which is represented by a temporal sequence of Burgers vortices of different
circulation, strain, and orientation. Such a model is motivated by the fact that
Navier-Stokes turbulence is known to create strong vortex filaments which tend to
dominate the time evolution of Lagrangian particles. This view is supported, e.g.,
by the recent work of Biferale and coworkers \cite{biferale04arx, biferale05pof} who detect
so-called ``vortex trapping events'' of tracer particles in their direct numerical
simulations. Hence we are led to consider the path of a Lagrangian particle in
the field of a single Burgers vortex. After a certain lifetime $T_l$, this
vortex decays and is replaced by another vortex which differs in circulation
$\Gamma$, strain-parameter $a$, and orientation. Then, the process starts anew.
Making suitable assumptions about the vortices' statistical properties, we will
be able to determine the statistics of the velocity increments in the framework of our model this way. In
particular, we will be able to investigate the origin and nature of small-scale
intermittency both numerically and analytically.

The remainder of the present article is structured as follows. Starting from the exact solution of a particle trajectory in a Burgers vortex, we will investigate the functional structure of the probability density function (pdf) of the velocity increment. We then will present results of a Lagrangian particle evolving through a temporal sequence of trapping events. There especially the evolution of the pdf's of the velocity increments is considered.

\section{Lagrangian particle in a single Burgers vortex}
The velocity field of a
single Burgers vortex with circulation $\Gamma$ in a strain field 
${\bf u}_{st}({\bf x},t)=[-\frac{a}{2} x, -\frac{a}{2} y, a z]$
is given by
\begin{equation}
{\boldsymbol u}({\boldsymbol x},t)={\boldsymbol u}_{st}({\boldsymbol x},t)+{\boldsymbol e_z} \times {\boldsymbol e}_r
\frac{\Gamma}{2 \pi r}
[1-e^{-ar^2/(4 \nu)}]
\end{equation}
where $\nu$ denotes the kinematic viscosity of the fluid, $\Gamma$ denotes
the circulation of the vortex and $r=\sqrt{x^2+y^2}$. $\boldsymbol e_z$ and $\boldsymbol e_r$ denote the unit vectors in $z$- and radial direction, respectively.
Switching now to the Lagrangian frame, the evolution of a Lagrangian particle starting from an initial position $\boldsymbol x_0$ is determined by
\begin{eqnarray}
  \frac{d \boldsymbol X}{d t}(\boldsymbol x_0,t)&=&\left[\boldsymbol u(\boldsymbol x,t)\right]_{\boldsymbol x=\boldsymbol X(\boldsymbol x_0,t)}\\
  \frac{d^2 \boldsymbol X}{d t^2}(\boldsymbol x_0,t)&=&\left[-\nabla p(\boldsymbol x,t)+\nu \Delta \boldsymbol u(\boldsymbol x,t)\right]_{\boldsymbol x=\boldsymbol X(\boldsymbol x_0,t)}. 
\end{eqnarray}
Due to the axial symmetry of the problem, we seek for solutions of the form
\begin{equation}
  \boldsymbol x(t)=r(t)\,\boldsymbol e_{r}(\varphi(t)) + z(t)\, \boldsymbol e_z,
\end{equation}
where $r(t)$, $\varphi(t)$ and $z(t)$ have to obey the following set of differential equations:
\begin{eqnarray}
  \dot r&=&-\frac{a}{2}r \\
  \dot{\varphi}&=&\frac{1}{r}u_{\varphi}(r) \\
  \dot z&=&az,
\end{eqnarray}
with $u_{\varphi}=\frac{\Gamma}{2 \pi r}
[1-e^{-ar^2/(4 \nu)}]$. The solutions for two of these components obviously read
\begin{eqnarray}
  z(t)&=&z_0\,e^{at} \\
  r(t)&=&r_0 e^{-\frac{a}{2}t}.
\end{eqnarray}
So the influence of the $z$-dynamics increases exponentially. Since we want to focus on the importance of the oscillatory motion around the axis of the vortex, we have to study the case of low straining, i.e. the factor $at$ has to be small compared to the circulation $\Gamma$. Plugging the solution for $r(t)$ into the ODE for the azimuthal component one obtains 
\begin{equation}
    \dot \varphi(t) = \frac{\Gamma}{2\pi r_0^2}e^{at}\left( 1-e^{-\frac{a}{4\nu} r_0^2e^{-at}} \right).
\end{equation}
The solution of this ODE is thereby reduced to an integration. 
However this integral cannot be solved explicitly, and hence we want to
restrict our following calculations to the limiting cases where 
the particle is far away from the vortex`s axis or near the core.
For the farfield case the point-vortex approximation holds, 
leading to an angular velocity of
\begin{equation}
  \dot \varphi(t)=\frac{\Gamma}{2\pi r_0^2}e^{at}.
\end{equation}
The angular velocity of the particle decays like $\frac{1}{r_0^2}$ leading 
to a differential rotation. In case of low straining the exponential can be approximated by unity. 
In this approximation the solution is
\begin{equation}
  \varphi(t)=\frac{\Gamma}{2\pi r_0^2}t=\omega(\Gamma, r_0)\, t.
\end{equation}
Note that due to this approximation the angular velocity becomes 
independent of the strain-parameter $a$.\\
The second limiting case is where the particle is near the viscous core 
of the vortex. In this case we can expand the exponential into a series. 
The first order approximation then is
\begin{equation}
  \dot \varphi(t)=\frac{\Gamma a}{8\pi\nu},
\end{equation}
which leads to
\begin{equation}
  \varphi(t)=\frac{\Gamma a}{8\pi\nu}\,t = \omega(\Gamma,a)\, t.
\end{equation}
This means that near the viscous core of the vortex the motion 
is given by a rigid body rotation determined by the parameters $a$, $\Gamma$ and $\nu$.
In order to clarify the functional structure of the following results for the velocity increment distribution we will first calculate the simple case, where only one vortex is involved. \\
Statistical observations on turbulence often focus on the velocity increments $v_x(\tau)$ with a time lag $\tau$ (see for example \cite{mordant04njp}) and on the turbulent acceleration \cite{aringazin04pre}. For the case of small $\tau$ the evolution of a particle in a turbulent flow is dominated by the nearest vortex filament. So the statistics for small $\tau$ can be derived directly from the dynamical equations of a single vortex.\\
Hence we have to write down the discussed solution in cartesian coordinates. The $x$-component of the position of the Lagrangian particle is given by
\begin{equation}
  x(t)=r_0\,e^{-\frac{a}{2}t}\cos{\varphi(t)}.
\end{equation}
Intermittency in the velocity signal is often said to be caused by the strong accelerations of a vortex filament \cite{lee05pre}. These accelerations originate from the oscillation of the particle round the axis of the vortex \cite{reynolds05njp}. We now neglect the radial transport of the particle in order to focus on these rotational accelerations. This approximation holds whenever the product $at$ is small compared to the vorticity determined by $\Gamma$. As we want to focus on the impact of strong vortex filaments on the statistics this is a reasonable approximation. Neglecting the straining the $x$-component of the velocity reads
\begin{equation}
  u_x(t)=-r_0\,\omega\,\sin{\omega t},  
\end{equation}
with $\omega$ being the oscillation frequency given by the two limiting cases discussed above. This leads to a simple equation for the velocity increments,
\begin{eqnarray}
  v_x(\tau)&=&u_x(t+\tau)-u_x(t) \nonumber \\
  &=&-r_0\,\omega\,\left( \sin\omega(t+\tau)-\sin\omega t \right).
\end{eqnarray}
Some simple trigonometric relations turn this expression to
\begin{eqnarray}
  v_x(\tau)&=&-2r_0\,\omega\,\left|\sin\frac{\omega\tau}{2}\right|\,\sin{(\omega t+\psi)} \nonumber \\
   &\equiv& A\,\sin{(\omega t+\psi)},
\end{eqnarray}
with a phase $\psi$ which depends on the initial conditions. Note that the velocity increments oscillate like the velocity components themselves, being modulated by an amplitude which depends on the parameters of the vortex as well as on the time lag $\tau$. The next aim is to deduce the corresponding probability density function. Instead of taking a time average, we average over the phase $\alpha=\omega t+\psi$, which can be assumed to be uniformly distributed,
\begin{eqnarray}
 f(v_x)&=&\frac{1}{2\pi}\int_{0}^{2\pi} \mathrm{d}\alpha\, \delta(v_x-A\sin\alpha) \nonumber \\
 &=&  \frac{1}{2\pi}\int_{0}^{2\pi} \mathrm{d}\alpha\, \frac{1}{|A\cos\alpha|} \, \delta(\alpha-\arcsin\frac{v_x}{A}).
\end{eqnarray}

Whenever $v_x<A$ the solution reads
\begin{eqnarray}
  f(v_x)&=&\frac{1}{\pi\sqrt{A^2-v_x^2}}\nonumber \\ 
  &=&\frac{1}{\pi\sqrt{4r_0^2\omega^2\sin^2(\frac{\omega\tau}{2})-v_x^2}}.
\end{eqnarray}
The dependence on the parameters $\Gamma$ and $a$ is absorbed into the frequency $\omega$. 
Noting that the violation of $v_x<A$ leads to imaginary values of the probability, the restriction becomes redundant when expressing the solution as 
\begin{eqnarray}
  f(v_x)&=&\Re\left( \frac{1}{\pi\sqrt{A^2-v_x^2}}\right) \nonumber \\ 
  &=&\Re\left(\frac{1}{\pi\sqrt{4 r_0^2\omega^2\sin^2(\frac{\omega\tau}{2})-v_x^2}}\right).
\end{eqnarray}
In order to obtain our result we have to average over the vorticity parameter $\Gamma$, the strain-parameter $a$ and the initial position of the particle, $r_0$ respectively. Let $f(\Gamma)$, $g(a)$ and $h(r_0)$ be the corresponding distributions of the parameters. (This notation implies statistical independence, a restriction, which has not to be made). The probability distribution of the velocity increments is then given by
\begin{equation}\label{p(vx)}
  p(v_x)=\Re\left(\int\frac{f(\Gamma)g(a)h(r_0)\,\,\mathrm{d}\Gamma \,\mathrm{d} a \,r_0\mathrm{d} r_0}{\pi\sqrt{4 r_0^2\omega^2\sin^2(\frac{\omega\tau}{2})-v_x^2}}\right).
\end{equation}
This expression explicitly reveals the dependence of the probability density functions on the distribution of the vorticity parameter $\Gamma$. Thereby a physical connection between the velocity increment distribution and the physical parameters of the flow is established.\\
The structure of the functional form of the pdf reveals that the shape of the pdf is given by a superposition of functions of the form $\frac{1}{\sqrt{A^2-x^2}}$ weighted by factors from the physical parameter distributions. The complex interplay of the parameter distribution then leads to the fat-tail pdf's, as observed in the following.\\

\begin{figure}
   \includegraphics[width=0.32\textwidth]{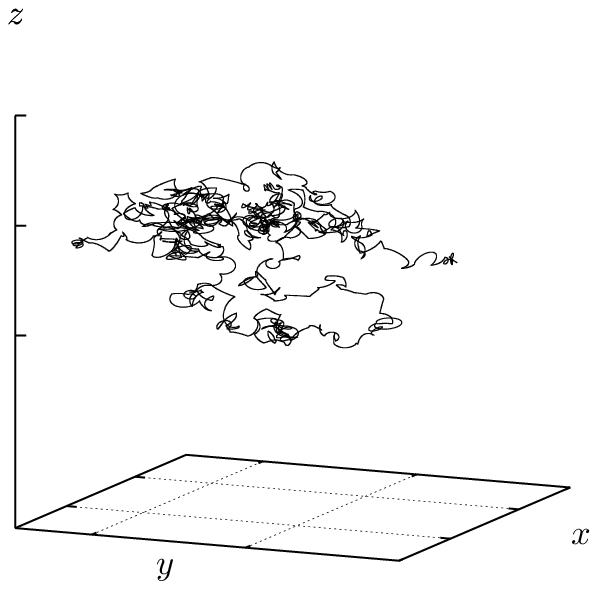}
   \includegraphics[width=0.4\textwidth]{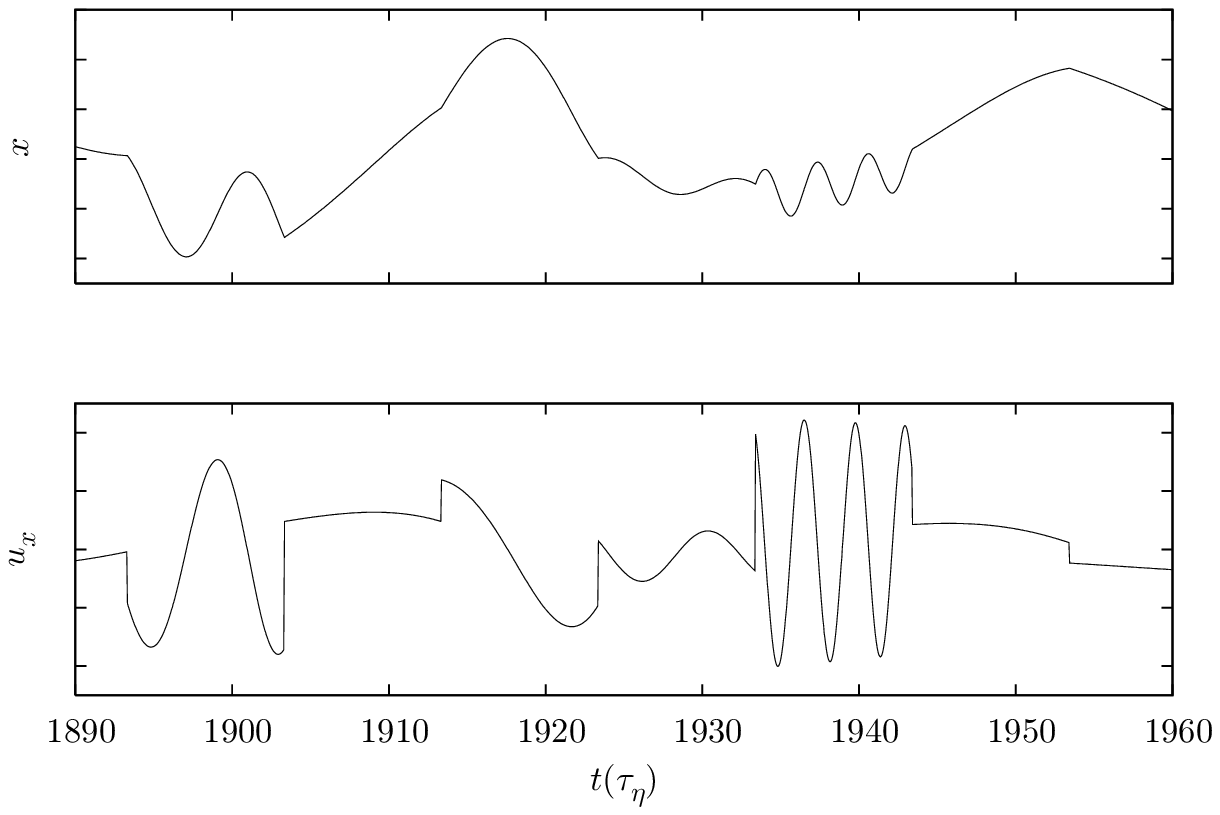}  
   \caption{\small Trajectory of a Lagrangian particle. $t$ is given in multiples of $\tau_{\eta}$.}
   \label{trajectory}
\end{figure}

\section{Lagrangian particle in a sequence of vortices}

\begin{figure}
   \includegraphics[width=0.4\textwidth]{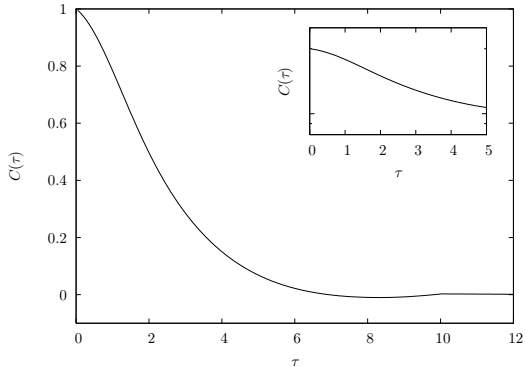}
   \caption{\small Autocorrelation function of the velocity $u_x$, $\tau$ is given in multiples of $\tau_{\eta}$. The inlet shows a log-plot of the vertically shifted autocorrelation function.}
   \label{autocorrelation}
\end{figure}

As discussed above, we model the evolution of a Lagrangian particle in a turbulent flow by a temporal sequence of randomly oriented coherent structures. In particular a particle is exposed to the velocity field of a single vortex for a given lifetime $T_l$. In our model $T_l$ is given by the typical lifetime of an eddy in the dissipative range. According to \cite{biferale04arx}, this lifetime is suggested to be of the order $10\tau_{\eta}$ ($\tau_{\eta}=(\varepsilon/\nu)^{\frac{1}{2}}$ denotes the Kolmogorov timescale). In real turbulent flows viscosity or vortex merging processes then force this vortex to decay. This is crudely modelled by simply turning the vortex off. Subsequently another vortex is switched on at a fixed distance. This distance is given by the typical spatial density of strong vortex filaments in a turbulent flow (see for example \cite{davidson04book}, chapter $7.3$). The orientation of the vortex is chosen randomly. As a result the statistics become isotropic.\\
In order to render the model more physical, some properties of turbulent flows have to be taken into account. Special interest has to be put on the statistical properties of the strain-parameter $a$ and the circulation $\Gamma$. A hint on the distribution of $\Gamma$ is given by the authors of \cite{jimenez93jfm}, who measured the distribution of the vortices' strengths directly from numerical simulations. In our model this is taken into account by modelling the distribution of $\Gamma$ with a log-normal distribution. Also the strain-parameter $a$ is chosen randomly. This means, that in our model vortices of different radii appear, as the strain-parameter $a$ directly affects the radius of the Burgers vortex. However, this distribution cannot be equally distributed, because this would cause vortices of any possible size to appear. Therefore a peaked distribution is needed. The expectation value of this distribution is related to the typical size of the smallest vortex filaments. We choose this distribution to be Gaussian with standard deviation $\sigma_a=1$. This choice is somewhat arbitrary, but it was ensured that the numerical results vary only little over a wide range of possible values of $\sigma_a$. The expectation value of the distribution can be derived from the definition of the radius of the Burgers vortex. This typical scale is defined as $ r_B=\left(  \frac{4\nu}{a}  \right)^{\frac{1}{2}}$. If we demand $r_B$ to be of the order of $10\eta$, a value which conforms to the findings of \cite{mouri03pre}, the expectation value of the strain-parameter $a$ is given as $ a=\frac{1}{25}\left(  \frac{\varepsilon}{\nu}  \right)^{\frac{1}{2}}=\frac{1}{25}\frac{1}{\tau_{\eta}}$, where $\varepsilon$ denotes the energy dissipation of the system. Note that by this relation a connection between typical scales of the velocity field of the Burgers vortex and physical properties of the flow is established. Moreover, the straining thereby is related to the Kolmogorov time scale $\tau_{\eta}$. A typical value of the circulation has to be determined in the same manner. We start with the well-known relation for the Reynolds number on the dissipative scale of the flow, $Re_{diss}=\frac{\eta u_{\eta}}{\nu}\approx 1$. The typical velocity $u_{\eta}$ can be approximated by the velocity field of the Burgers vortex on the scale $\eta$,
\begin{equation}
  u_{\eta}=u_{\varphi}(\eta)=\frac{1}{2\pi\eta}\left(  1-e^{-\frac{a\eta^2}{4\nu}}  \right) \approx \frac{\Gamma a \eta}{8\pi\nu}.
\end{equation}
This yields $ \Gamma=\frac{8\pi}{a}\left(  \epsilon \nu  \right)^\frac{1}{2}=200\pi\nu$, which determines a typical value for $\Gamma$. The distribution of $\Gamma$ is assumed to be log-normal according to the observations of \cite{jimenez93jfm}.\\
Now all properties of the model are specified and we are able to perform a numerical simulation.\\
Figure \ref{trajectory} shows a typical particle path. It reveals that the path is composed of a sequence of vortex trapping events, each of them characterized by a different circulation $\Gamma$. Events with a strong circulation can easily be identified by their rapid oscillation of the $x$-component, whereas events with small values of $\Gamma$ do not cause the particle to perform a complete oscillation. 
The switching between vortices renders the velocities discontinuously. This discontinuity can be eliminated using a low pass filter. A lowpass filtering of the velocity signal with a Butterworth filter was performed and the resulting statistics was compared to the unfiltered one. It was ensured that this effect has no significant impact on the statistics.\\
\begin{figure}
   \includegraphics[width=0.4\textwidth]{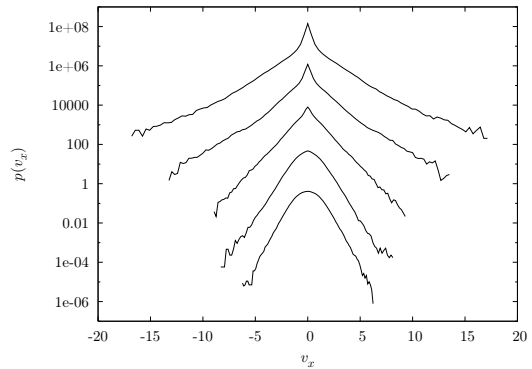}
   \caption{ Logarithmic plot of the pdf's of the velocity increments. Time lags $\tau \epsilon \{(0.98,1.97,3.94,7.87,15.74)\tau_{\eta}\}$. The pdf's are normalized to $\sigma=1$ and shifted vertically for viewing convenience.}
   \label{increments}
\end{figure}
\begin{figure}
   \includegraphics[width=0.4\textwidth]{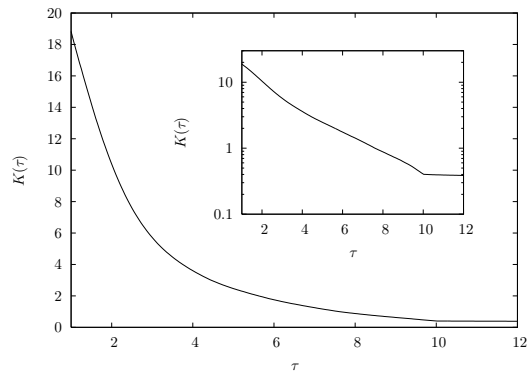}
   \caption{\small Kurtosis $K(\tau)=\frac{\langle v_x(\tau)^4 \rangle}{\langle v_x(\tau)^2 \rangle^2}-3$ for $p(v_x)$. The inlet shows a log-plot of $K(\tau)$. The kurtosis shows an almost exponential decay.}
   \label{kurtosis}
\end{figure}
One of the central results of our model is depicted in figure \ref{increments} which shows the probability distribution of the velocity increments. These exhibit a similar behavior to the velocity increment distributions observed in experiments or direct numerical simulation. For small time lags $\tau$ the pdf's show fat tails, whereas a nearly Gaussian shape arises for large time lags $\tau$. The physical explanation is straight forward; while the functional form for the short time lags is almost solely determined by a single vortex, the transition to a more Gaussian functional form results from an increasing statistical independence as more vortices are involved. The transition from one functional form to the other is characterized in more detail by the kurtosis, which is shown in figure \ref{kurtosis}. This figure clearly indicates the transition from a highly intermittent distribution to a more Gaussian one.
Since the subsequent trapping events are statistically independent, the velocity signal has to decorrelate as a function of the time delay $\tau$. We define the velocity autocorrelation function as
\begin{equation}
  C(\tau)=\frac{\langle u_x(t+\tau)\,u_x(t) \rangle}{\langle u_x^2 \rangle}.
\end{equation}
Figure \ref{autocorrelation} confirms this view; the correlation decreases almost exponentially. The slight anticorrelations might tribute to the events with strong circulations, where the particle is able to perform one or more complete oscillation. The velocity signal decorrelates after a time of $\tau=10 \tau_{\eta}$, which is the lifetime of our vortices. A second effect which cannot be clearly separated is dephasing. As the particle moves inward it constantly alters its angular velocity. It can be shown easily that this also causes the autocorrelation function to decrease. \\

\section{summary}
To conclude, we modelled the particle evolution in a turbulent flow as a temporal sequence of Burgers vortices. We are able to calculate the path of a Lagrangian particle in an isolated vortex. From this point we derived the functional form of the velocity increment pdf. This calculation reveals the dependence of the structure of the velocity increment distribution on the physical parameters of the vortex filament like strain, circulation etc. \\
A numerical implementation of the model was used to investigate some typical Lagrangian properties of turbulent flows. As a central Lagrangian observable we focused on the velocity increment distribution finding that our model qualitatively resembles the velocity increment statistics in real turbulent flows. A further investigation of the corresponding kurtosis clearly indicated intermittent characteristics. \\
Additionally the velocity autocorrelation function was investigated revealing an almost exponential decrease and even slight anticorrelations. Consistent with the framework of the model is the fact that the autocorrelation function vanishes for times longer than the lifetime of a single vortex. \\
The physical interpretation of the observed results is quite straightforward. A Lagrangian particle encountering subsequent trapping events shows intermittent velocity increment distributions accountable to a proper statistical ensemble of Burgers vortices. For small time lags the fat-tail statistics originate from the superposition of the dynamics in a single vortex. The transition to a more Gaussian behavior is caused by an increasing mixing of two subsequent vortex trapping events which are statistically independent. This observation is supported by the functional form of the velocity autocorrelation function. \\
This model qualitatively reproduces some of the typical Lagrangian statistics. It is important to note that this is achieved by a temporal composition of exact solutions of the Navier-Stokes equation. This is in contrast to many other models, which apply stochastic equations for the particle evolution. Due to the simplicity of the model our results can be interpreted in physical terms and shed light on the connection between dynamical aspects in turbulent flows and corresponding statistical properties.



\end{document}